\documentclass[structabstract]{aa}
\usepackage{amsmath}
\usepackage{graphicx}
\usepackage{natbib}
\usepackage{color}
\usepackage{txfonts}
\def\be{\begin{equation}}
\def\ee{\end{equation}}
\def\bdm{\begin{displaymath}}
\def\edm{\end{displaymath}}

\begin{document}
   \title{On the interpretation and applicability of $\kappa$-distributions}
   \titlerunning{On $\kappa$-distributions}
   \author{M.\ Lazar \inst{1,2}
   \fnmsep\thanks{\email{mlazar@tp4.rub.de}},
    H.\ Fichtner \inst{2,3} and P.\ H.\ Yoon \inst{4,5}}
         \authorrunning{M.\ Lazar, H.\ Fichtner and P.\ Yoon}
   \institute{
   $^1$ School of Mathematics and Statistics, University of St Andrews, 
	St Andrews, Fife, KY16 9SS, U.K.\\
   $^2$ Institut f\"ur Theoretische Physik, Lehrstuhl IV: Weltraum- und
        Astrophysik, Ruhr-Universit\"at Bochum, D-44780 Bochum, Germany\\
   $^3$ Research Department of Complex Plasmas, Ruhr-Universit\"at Bochum,
        D-44780 Bochum, Germany\\
   $^4$ Institute for Physical Science and Technology, University of
        Maryland, College Park, USA\\
   $^5$ School of Space Research, Kyung Hee University, Korea}
\date{Received October, 2015; accepted , }

   \abstract
   {The generally accepted representation of $\kappa$-distributions
   in space plasma physics allows for two different
   alternatives, namely assuming either the
   temperature or the thermal velocity to be $\kappa$-independent.}
   {The present paper aims to clarify the issue concerning which
   of the two possible choices and the related physical interpretation
   is the correct one.}
   {A quantitative comparison of the consequences of the use of
   both distributions for specific physical systems leads to
   their correct interpretation.}
   {It is found that both alternatives can be realized, but are
   valid for principally different physical systems.}
   {The investigation demonstrates that, before employing one
   of the two alternatives, one should be conscious about the nature
   of the physical system one intends to describe, otherwise one
   would possibly obtain unphysical results.}


   \maketitle
%

\section{Introduction}
The $\kappa$-distributions are useful tools for quantitative
treatment of non-thermal space and astrophysical plasmas
\citep[e.g.,][and references therein]{Pierrard-Lazar-2010, Livadiotis-McComas-2013, Fahr-etal-2014}.
After their heuristic first definition almost 50~years ago they have not only
been used in innumerable applications, but various authors have successfully
derived $\kappa$-distributions more rigorously for specific physical
system, such as \citet{Hasegawa-etal-1985}, who considered a
plasma in a prescribed suprathermal radiation field, or
\citet{Ma-Summers-1998}, who assumed the presence of prescribed
stationary whistler turbulence. More recent example may be
\citet{Yoon-2014}, who self-consistently solved the problem of
an electron distribution that is in a dynamic equilibrium with
electrostatic Langmuir turbulence. Other authors even attempted
to motivate the physical significance of $\kappa$-distributions
from fundamental principles, such as \citet{Tsallis-1988}, who
considered the generalized version of the Renyi entropy, or
\citet{Treumann-Jaroschek-2008}, who constructed a statistical
mechanical theory of such power-law distributions via generalizing
Gibbsian theory. Despite these theoretical foundations, there
exists as yet no generally accepted unique interpretation of
$\kappa$-distributions
\citep[see][and references therein]{Livadiotis-2015}.
As discussed recently in \citet{Lazar-etal-2015}, one can
rather distinguish two principally different alternatives.

The first choice dates back to the original idea for the definition
of $\kappa$-distributions, which first appeared in printed form in
the paper by \citet{Vasyliunas-1968} but can be traced back to
Stanislav Olbert, as acknowledged by the author himself. When Olbert
introduced it in one of his own papers published a few
month later in the same year, he motivated his definition of
$\kappa$-distributions in the context of magnetospheric electron
spectral measurements as follows \citep{Olbert-1968}: {\it ``[...]
the electron speed distribution [...] is of the form}
\begin{equation}
f_e v^2 dv = const\,\left(1+\dfrac{v^2}{\kappa w_0^2}\right)^{-\kappa-1} \, v^2 dv
\label{olbert_vasyliunas}\end{equation}
{\it where $v$ is the actual speed, $w_0$ is the most probable speed
of electrons, and $\kappa$ is a `free' parameter whose value is a
measure of the departure of the distribution from its Maxwellian
character (letting $\kappa$ approach infinity leads to the
Maxwellian distribution). We shall not go into the reasons for this
choice except to mention that it seems to be justifiable on the
basis of other independent observations.''} Evidently, with this
ad-hoc definition \citet{Olbert-1968} intended to heuristically
describe an enhanced fraction of suprathermal particles, as compared
to a Maxwellian distribution. Naturally, such suprathermal
$\kappa$-distribution is characterized by a higher
$\kappa$-dependent temperature.

Contrary to this expectation \citet{Livadiotis-2015} recently
offered a different view by stating: {\it ``The temperature acquires
a physical meaning as soon as the Maxwell's kinetic definition
coincides with the Clausius's thermodynamic definition [...]. This
is the actual temperature of a system; it is unique and independent
of the kappa index.''}

In order to have a $\kappa$-independent temperature, it is easy to
see (section~2) that one must consider the thermal velocity (called
$w_0$ in Olbert's definition) to be $\kappa$-dependent. A little
more subtle aspect is another consequence of this assumption, namely
that it not only implies an enhancement of the velocity distribution
(VDF) relative to the associated Maxwellian at higher velocities but
also the enhancement of the core population at very low velocities.

The obvious question that arises is: {\it Which of the two
interpretations is correct or can both be valid for different
physical systems?} The purpose of the present paper is to answer
this question. To this end we define the $\kappa$-distributions
explicitly in section~2, critically discuss their physical
implications in sections~3 and 4, and summarize our findings
in the concluding section~5.
\section{Definitions of the $\kappa$-distributions}

In most general from one can define bi-$\kappa$-distributions
in a magnetized plasma as follows
\citep{Lazar-etal-2015}:
\begin{eqnarray}
F_K (v_{\parallel}, v_{\perp}) & = & {1 \over \pi^{3/2}
\theta_{\perp}^2 \theta_{\parallel}} \, {\Gamma(\kappa +1) \over
\kappa^{3/2} \Gamma(\kappa -1/2)} \left(1 + {v_{\parallel}^2\over
\kappa \theta_{\parallel}^2 } + {v_{\perp}^2\over \kappa
\theta_{\perp}^2 }\right)^{-\kappa-1}
\nonumber \\
& = & \left[m \over \pi k_B (2\kappa-3)\right]^{3/2}
{1 \over T^K_\perp \sqrt{T^K_\parallel}}
{\Gamma(\kappa +1) \over \Gamma(\kappa -1/2)}
\nonumber\\
& \times & \left[1+ {m \over k_B (2\kappa-3)} \left(
{v_{\parallel}^2\over T^K_{\parallel} } - {v_{\perp}^2\over
T^K_{\perp} }\right) \right],
\label{kappas}
\end{eqnarray}
where $v_{\|}$ and $v_{\perp}$ denote particle velocity parallel and
perpendicular w.r.t.\ a large-scale magnetic field, $T_{\|,\perp}$
and $\theta_{\|,\perp}$ the corresponding temperatures and thermal
velocities, which are related by
\begin{eqnarray}
T_{\parallel}^K & = & {m \over k_B} \int d{\bf v} v_{\parallel}^2
F_K (v_{\parallel}, v_{\perp}) = {m \over 2 k_B} \, {2 \kappa
\over 2 \kappa -3} \theta_{\parallel}^2
\label{tpara}\\
T_{\perp}^K & = & {m \over 2 k_B} \int d{\bf v} v_{\perp}^2 F_K
(v_{\parallel}, v_{\perp}) =  {m \over 2 k_B} \, {2 \kappa \over 2
\kappa -3} \theta_{\perp}^2.
\label{tperp}
\end{eqnarray}
In the above $m$ is particle mass, $k_B$ the Boltzmann constant,
$\Gamma$ is the Gamma function, and $\kappa \in (3/2,\infty]$.

As already pointed out in \citet{Lazar-etal-2015}, despite this
general formulation of the bi-$\kappa$-distributions, the
interpretation for the temperatures can be ambiguous, as it can be
understood and interpreted in two alternative ways:
\begin{description}
    \item[(A)] The temperatures of the bi-$\kappa$-distributions and
    of the associated bi-Maxwellian are identical,
    i.e.\ $T^K_{\parallel,\perp} = T^M_{\parallel,\perp}$.
    This implies that thermal velocities are $\kappa$-dependent via
        \begin{equation}
        \theta_{\parallel,\perp}
        = \sqrt{\left(1- {3 \over 2 \kappa}\right)
        \,\frac{2 k_B T^M_{\parallel,\perp}}{m}}
        \label{theta_A}\end{equation}
    This corresponds to the alternative interpretation
    advocated by \citet{Livadiotis-2015}.
    \\
    \item[(B)] Thermal velocities of bi-$\kappa$-distributions
    and of the associated bi-Maxwellian are identical, i.e.\
        \begin{equation}
        \theta_{\parallel,\perp}=
        \sqrt{\frac{2 k_B T^M_{\parallel,\perp}}{m}}.
        \label{theta_B}\end{equation}
    This implies that the temperatures are $\kappa$-dependent via
        \begin{eqnarray}
        T^K_{\parallel,\perp} = {2\kappa \over 2 \kappa - 3}
        \, {m \theta_{\parallel,\perp}^2 \over 2k_B}
        = {2\kappa \over 2 \kappa - 3}
        \, T^M_{\parallel,\perp} > T^M_{\parallel,\perp}
        \end{eqnarray}
     This corresponds to the original definition by \citet{Olbert-1968}.
\end{description}
\begin{figure}[h!]
   \begin{center}
   \includegraphics[width=0.85\columnwidth]{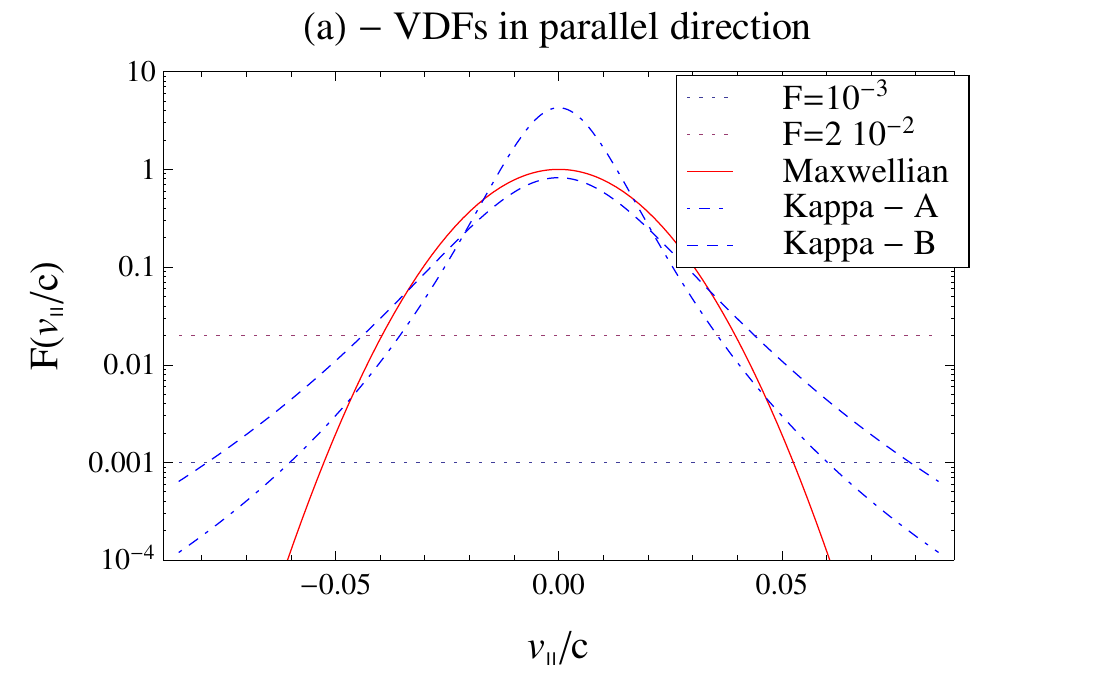} \\
   \hspace*{-0.85cm}
   \includegraphics[width=0.77\columnwidth]{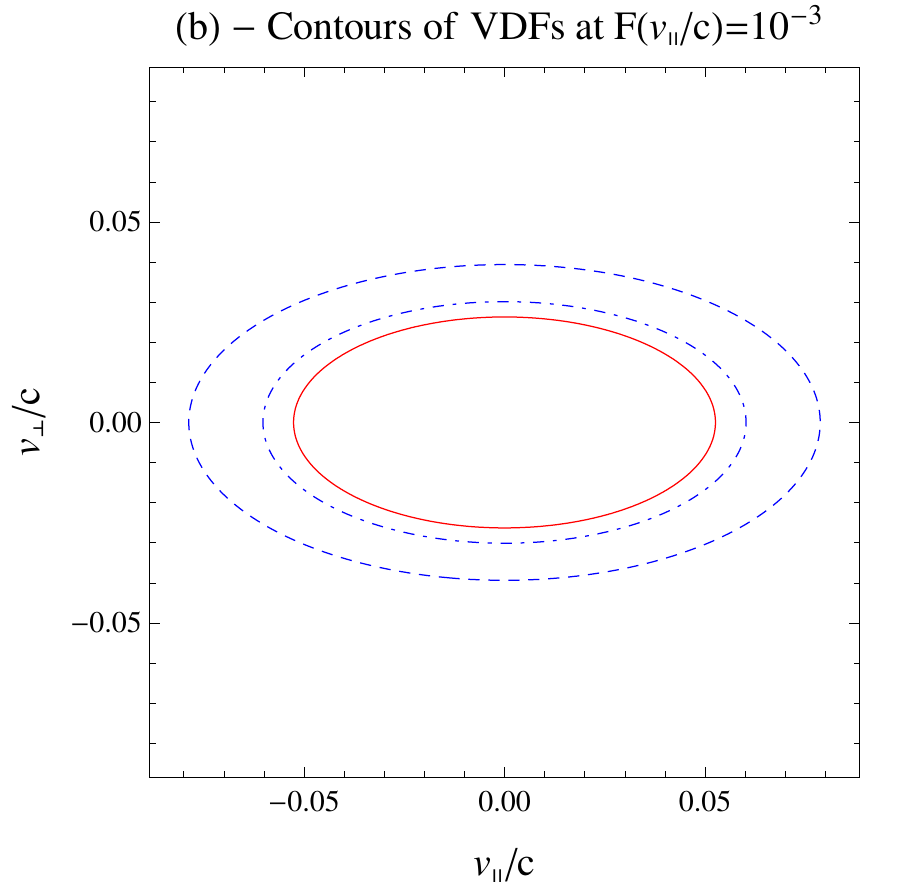}\\
   \hspace*{-0.85cm}
   \includegraphics[width=0.77\columnwidth]{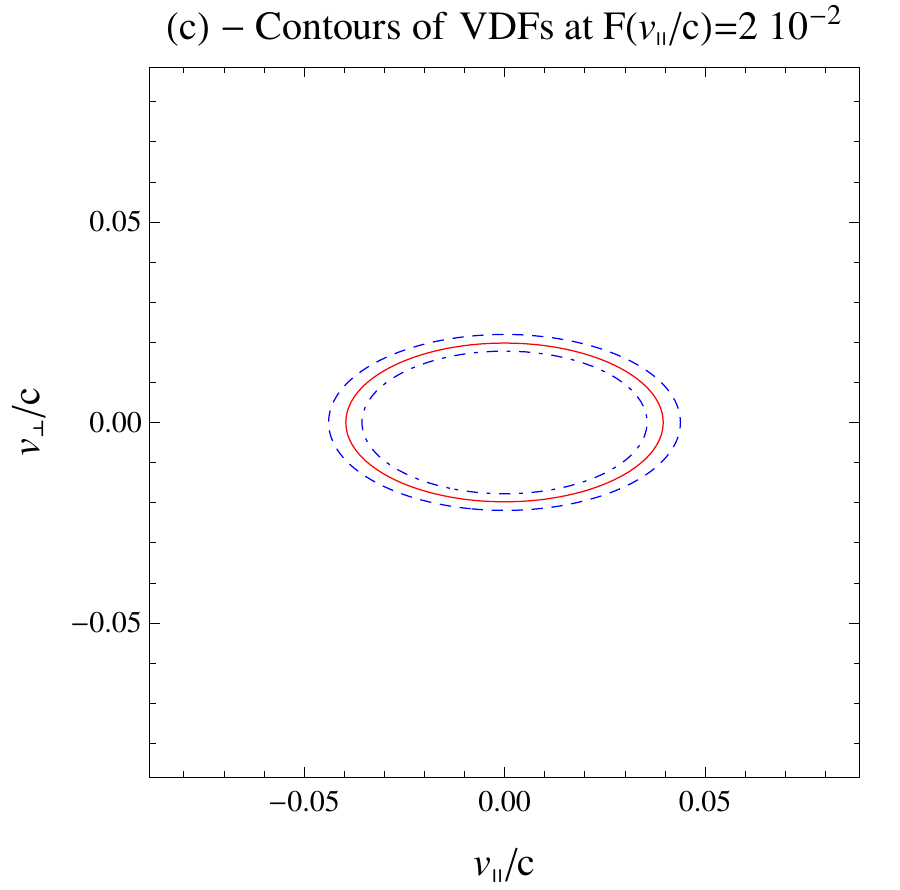}
   \end{center}
   \caption{The two alternative bi-$\kappa$-distributions
   for $\kappa=9/4$ and the associated Maxwellian model:
   Panel (a) gives their parallel parts and panels (b) and (c) show
   the contours of the full VDFs at the levels (dotted lines) indicated in panel (a).
   Evidently both bi-$\kappa$-distributions exhibit enhanced
   tails relative to the Maxwellian but Kappa-A has, in addition,
   also an enhanced core.  } \label{fig1}
\end{figure}
In the following we refer to these alternatives as `Kappa~A' and
'Kappa-B', respectively. These $\kappa$-distributions differ in a
crucial way, as illustrated in Fig.~\ref{fig1}, where (a) the
parallel part of the two $\kappa$-distributions for $\kappa=9/4$ is
shown along with the associated Maxwellian, and where (b) contour
plots of the VDFs at the level indicated in panel (a) are given.
As is evident from panel (a), per construction, both
$\kappa$-distributions are enhanced at higher velocities relative to
the Maxwellian. However, interestingly, Kappa-A additionally exhibit
an increased core population. Consequently, unavoidably, a question
arises, namely {\it Which of the two $\kappa$-distributions is the
correct one?} The answer should be found on the basis of
quantitative modeling and by considering the consequences of their
use for specific physical scenarios. This is discussed in the
following section.

\section{Comparison of the two $\kappa$-distributions}

\subsection{Non-Maxwellian plasmas due to reduced interactions}

One argument for the formation of enhanced suprathermal VDF tails,
for example in the solar wind, is the lack of collisions or,
more generally, due to insufficient interactions, which could
maintain a Maxwellian equilibrium. In this case, it is expected
that there should be comparatively more particles with higher
velocities, and less particles with lower velocities
\citep[e.g.,][]{Fichtner-Sreenivasan-1993}.
With a glance at the original purpose one would prefer
Kappa-B for such a scenario: \citet{Olbert-1968}
intended to describe a particle velocity distribution that has,
compared to a Maxwellian, an enhanced fraction of suprathermal
particles. Such an enhanced halo of the VDF must be expected to
form at the expense of its core population, i.e.\ one must
expect the modified distribution to be depleted at low velocities.

This expectation has been confirmed with the recent direct
comparison of using Kappa-A or Kappa-B versus
the associated Maxwellian in the studies of plasma waves,
and the discussion of the consequences thereof by
\citet{Lazar-etal-2015}. These authors investigated the
electromagnetic electron-cyclotron waves driven by
perpendicular temperature anisotropy, $T_{\perp}/T_{\|} > 1$,
on the basis of solutions of the corresponding dispersion relation
that can be cast into the form
\begin{equation}
A^K + {A^K \,(\omega - \Omega)+\Omega \over k \theta_{\parallel}} \;
Z_{\kappa} \left({\omega - \Omega \over k
\theta_{\parallel}}\right)
- {k^2c^2 \over \omega_p^2} - 1 =0,
\label{e14}
\end{equation}
with the temperature anisotropy $A^K=T^K_{\perp}/T^K_{\|}$, the complex
wave frequency $\omega(k) = \Re(\omega)(k) + i \Im(\omega)(k)$, the wave number $k$, the gyrofrequency $\Omega$,
the plasma frequency $\omega_p$, and the speed of light $c$. The function
\begin{equation}
Z_{\kappa}(z) = \frac{1}{(\pi\kappa)^{1/2}} {\Gamma
(\kappa) \over \Gamma \left(\kappa - 1/2\right)}
\int\limits_{-\infty}^{+\infty}\! {(1+x^2/\kappa)^{- \kappa } \over x - z}\, dx , \;\; \Im (f) > 0
\label{e15}
\end{equation}
is the modified $\kappa$-plasma dispersion function \citep{Lazar-etal-2008}.
For $\kappa\to \infty$ one recovers the dispersion relation for a Maxwellian
plasma with the standard plasma dispersion function \citep{Fried-Conte-1961}.

\begin{figure}[h!]
   \begin{center}
   \hspace*{0.45cm}
   \includegraphics[width=0.95\columnwidth]{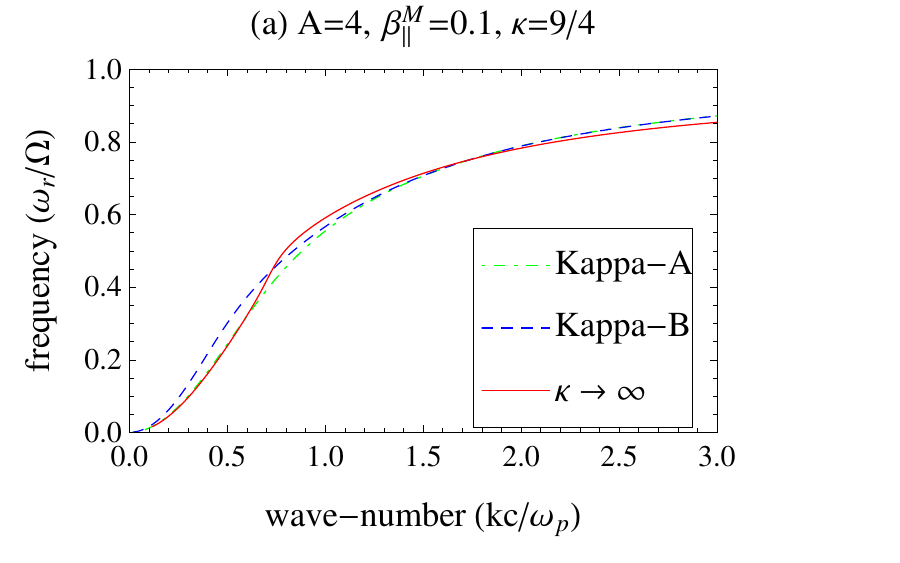}\\
   \includegraphics[width=0.95\columnwidth]{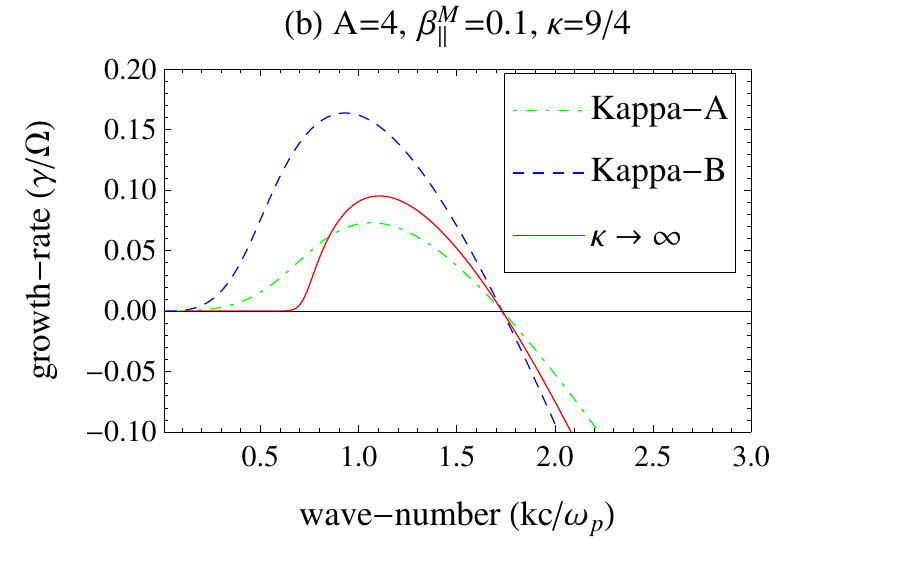}
   \end{center}
   \caption{The (a) dispersion curves $\Re(\omega)(k)$ and (b) growth rates
    $\Im(\omega)(k)$ (b) derived for a bi-Maxwellian (solid lines), a
    bi-Kappa-A (dot-dashed lines), and a bi-Kappa-B (dashed lines) for
    $A^{K,M} =4$, a plasma $\beta_{\|}^M = 0.1$ and $\kappa = 9/4$.}
\label{fig2}
\end{figure}

\citet{Lazar-etal-2015} demonstrates that, while all VDFs give very
similar dispersion curves, there are significant differences in the
growth rates for given anisotropic plasma conditions. One would
expect an enhanced fraction of suprathermal particles to increase
the growth rates systematically and monotonously, i.e.\ there should
be no wave-number interval with lower growth rates when compared to
the Maxwellian. This behavior is exactly exhibited by Kappa-B. In
contrast, the use of Kappa-A results in non-monotonously higher {\it
or} lower growth rates than the Maxwellian, an example is provided
with Fig.~\ref{fig2}.

Consequently, for a plasma scenario apparently envisaged by
\citet{Olbert-1968}, i.e.\ a VDF with an enhanced tail but
not an additionally enhanced core, the answer to the above
question is that Kappa-B is the correct choice.

\subsection{Non-Maxwellian plasmas due to specific
wave-particle interactions}

It has also been suggested that the high-velocity power-law tails
can form due to specific wave-particle interactions, e.g.\ due to
electromagnetic waves \citep{Hasegawa-etal-1985}, Whistler waves
\citep{Ma-Summers-1998}, fast-mode waves
\citep{Roberts-Miller-1998}, Alfv\'en waves
\citep{Leubner-2000}, or stochastic acceleration by
turbulence of arbitrary nature but characterized by a diffusion
coefficient with an inverse dependence of velocity
\citep{Bian-etal-2014}. In improvement of such test-particles
approaches, \citet{Yoon-2014} self-consistently solved the problem
of an (isotropic) electron distribution that is in a dynamic
equilibrium with electrostatic Langmuir turbulence.

To briefly overview Yoon's theory, the steady-state isotropic
electron VDF $F(v)$ in the presence of Langmuir turbulence
intensity $I_L(k)=E_k^2$ is given by
\begin{equation}
F(v)=C\exp\left(-\int\frac{mv}{4\pi^2}\frac{1}{{\cal J}(v)}\right).
\label{gen_F}
\end{equation}
with
\begin{equation}
{\cal J}(v)=\frac{1}{{\cal H}(v)}\int_{\omega_p/v}^\infty
I_L(k)\frac{dk}{k}\;\;,\;\;
{\cal H}(v)=\int_{\omega_p/v}^\infty\frac{dk}{k}.
\label{H_J}
\end{equation}
This solution is derived from the particle kinetic equation that
describes diffusion and friction (or drag) in velocity-space arising
from the spontaneously emitted electrostatic Langmuir fluctuations.
With ${\cal J}={\rm const}=k_BT_M/(4\pi^2)$,
where $T_M$ is the isotropic Maxwellian temperature, one
obtains the Maxwell distribution, $F_M(v)=C\exp\left(-mv^2/2k_BT_M\right)$.
However, \citet{Yoon-2014} assumed a generalized Kappa distribution,
\begin{equation}
F(v)=\frac{1}{(\pi\theta^2)^{3/2}}
\frac{\Gamma(\kappa+1)}{{\kappa'}^{3/2}\Gamma(\kappa-1/2)}
\frac{1}{(1+v^2/\kappa'\theta^2)^{\kappa+1}},
\label{gen_kappa}
\end{equation}
where it should be noted that, unlike the customary $\kappa$-model,
$\kappa'$ is generally allowed to be different from $\kappa$.
The effective temperature is given by
\begin{equation}
T=\int d{\bf v}\frac{mv^2}{3k_B}F
=\frac{m}{2k_B}\frac{2\kappa'}{2\kappa-3}\theta^2.
\label{T_theta}
\end{equation}
Note that the above definition is essentially the same
as (\ref{tpara}) and (\ref{tperp}), except that
(\ref{T_theta}) defines isotropic temperature and
that on the right-hand side of the last equality,
the numerator is given by $\kappa'$ instead of $\kappa$.

Upon comparing the assumed solution (\ref{gen_kappa})
and the formal steady-state solution (\ref{gen_F}), it quickly
becomes obvious that the reduced Langmuir fluctuation
spectrum ${\cal J}$ must be given by
\begin{equation}
{\cal J}(k)=\frac{m\theta^2}{8\pi^2}
\frac{\kappa'}{\kappa+1}\left(1
+\frac{\omega_p^2}{\kappa'k^2\theta^2}\right).
\label{J}
\end{equation}
It also follows from the definitions of ${\cal H}$
and ${\cal J}$ given by (\ref{H_J}) that the full
Langmuir intensity can be deduced as
\begin{equation}
I_L(k)=\frac{m\theta^2}{8\pi^2}\frac{\kappa'}{\kappa+1}
\left(1+\frac{\omega_p^2}{\kappa'k^2\theta^2}
\left[1+2{\cal H}(k)\right]\right).
\label{IL}
\end{equation}
Note that with ${\cal H} = 0$
\citep[see the discussion in][]{Yoon-2014}
Eqs.(\ref{J}) and (\ref{IL}) become identical.

\citet{Yoon-2014} subsequently demonstrated that the
solution ${\cal J}(k)$ for the reduced Langmuir fluctuation
spectrum is also, consistently, the steady-state solution of
the wave kinetic equation, when exclusively linear wave-particle
interactions are considered. Including the nonlinear terms in the
wave kinetic equation, \citet{Yoon-2014} re-derived the exact
solution for the full Langmuir intensity as
\begin{equation}
I_L(k)=\frac{k_BT_i}{4\pi^2}\left[1+\frac{4}{3}
\left(\kappa-\frac{3}{2}\right)
\frac{\omega_p^2}{\kappa'k^2\theta^2}\right],
\label{IL_alt}
\end{equation}
which must be identical to (\ref{IL}). Consequently,
it immediately follows that
\begin{eqnarray}
\kappa' &=& (\kappa+1)\frac{2k_BT_i}{m\theta^2}
=\left(\kappa-\frac{3}{2}\right)\frac{2k_BT}{m\theta^2},
\nonumber\\
\kappa &=& \frac{9}{4}+\frac{3{\cal H}}{2},
\label{kappaPrime_kappa}
\end{eqnarray}
which with ${\cal H} = 0$
\citep[see the discussion in][]{Yoon-2014} leads to $\kappa=9/4$.

Consequently, the self-consistent solution can be summarized by
a coupled set of solutions,
\begin{eqnarray}
F(v) &=& \frac{m_e^{3/2}}{(2\pi T_e)^{3/2}}
\frac{\Gamma(\kappa+1)}{(\kappa-3/2)^{3/2}
\Gamma(\kappa-1/2)}
\nonumber\\
&& \times\left(1+\frac{1}{\kappa-3/2}
\frac{m_ev^2}{2T_e}\right)^{-\kappa-1},
\nonumber\\
I_L(k) &=& \frac{k_BT_e}{4\pi^2}\frac{\kappa-3/2}
{\kappa+1}\left(1+\frac{1}{\kappa-3/2}
\frac{2\pi ne^2}{k^2T_e}\right),
\nonumber\\
\kappa &=& \frac{9}{4},\qquad\frac{T_i}{T_e}
=\frac{\kappa-3/2}{\kappa+1},
\label{yoon_sol}\end{eqnarray}
where $\kappa'$ no longer appears. Clearly, the
electron VDF is of the type Kappa-A. A noteworthy
feature associated with the Langmuir intensity is
that the long-wavelength regime ($k\to 0$) is enhanced
over the Maxwellian limit, $I_L(k)=k_BT_e/(4\pi^2)$, while
for short wavelengths, the Langmuir fluctuation
spectrum decreases relative to the Maxwellian one.
It is the relation (\ref{kappaPrime_kappa}), particularly
the specific identification of $\kappa'=(2\kappa-3)k_BT/m\theta^2$,
which renders (\ref{T_theta}) into the $\theta^2$ vs $T$
relationship of the first type, which in turn,
led to the Kappa-A model.

%
%

As evidenced from figure 1, the Kappa-A distribution
self-consistently constructed by \citet{Yoon-2014}
exhibits not only an enhanced high-velocity tail but also
an enhanced core population. The latter enhancement can
only be understood if a process exist that keeps more
particles (relative to the Maxwellian) at low velocities.
In the final solution (\ref{yoon_sol}) this can be understood
in the context of the wave-particle resonance condition,
$\omega_p\simeq{\bf k}\cdot{\bf v}$ between the low-velocity
electrons and reduced Langmuir fluctuation spectrum
in the short-wavelength regime ($k\gg1$). The reduced Langmuir
intensity spectrum (relative to the Maxwellian case) leads
to an accumulation of low-velocity electrons near $v\sim0$,
as the wave-particle resonance becomes ineffective,
while for high-velocity particles the enhanced
Langmuir intensity near $k\sim 0$ leads to acceleration
and, thereby, to the formation of the power-law tail.

Consequently, one can state in general that, if a process exists
that keeps more particles (relative to a Maxwellian) at low
velocities, the answer to the above question is that Kappa-A
is the correct choice: The low-velocity enhancement balances the
high-velocity enhancement, so that the temperature is indeed independent
of the parameter $\kappa$.

\section{An alternative view: Two Maxwellian limits for a given $\kappa$-distribution}

\begin{figure}[h!]
\centering
   \includegraphics[width=0.85\columnwidth]{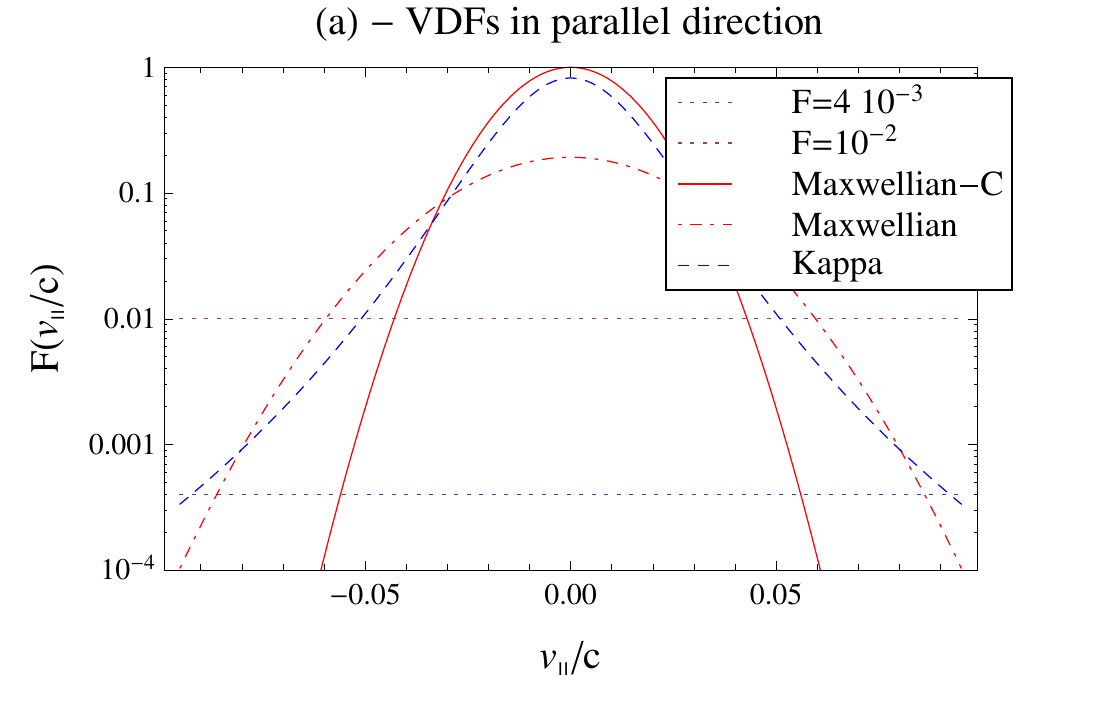} \\
   \hspace*{-0.8cm}
   \includegraphics[width=0.77\columnwidth]{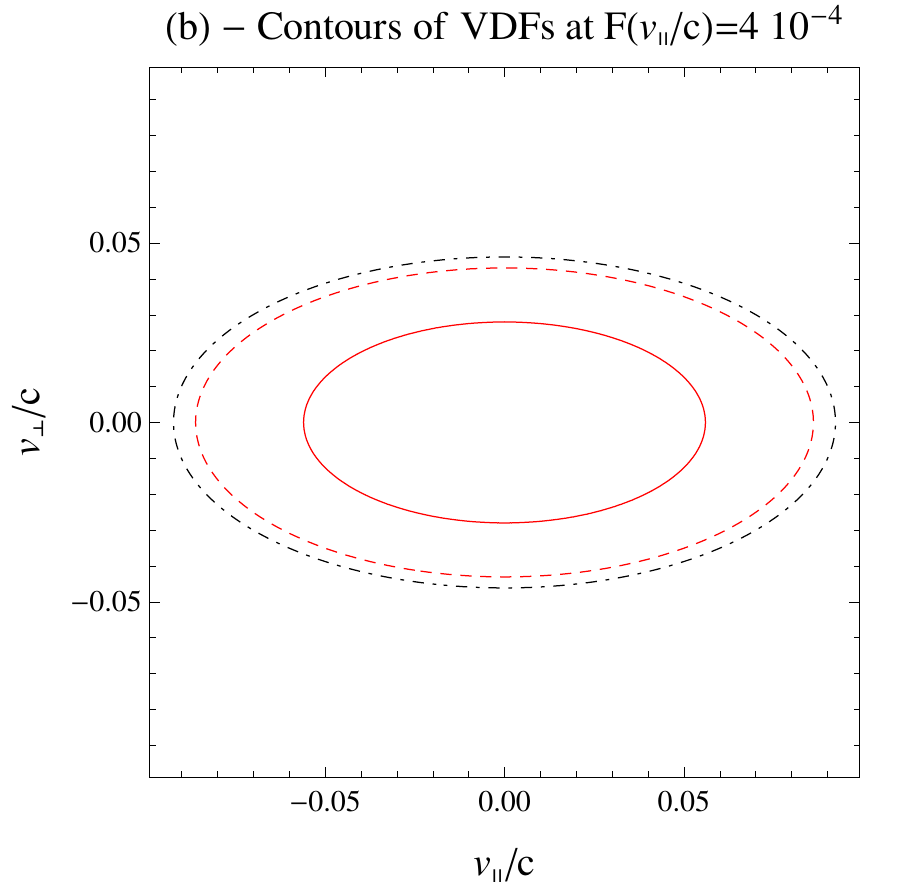} \\
   \hspace*{-0.8cm}
   \includegraphics[width=0.77\columnwidth]{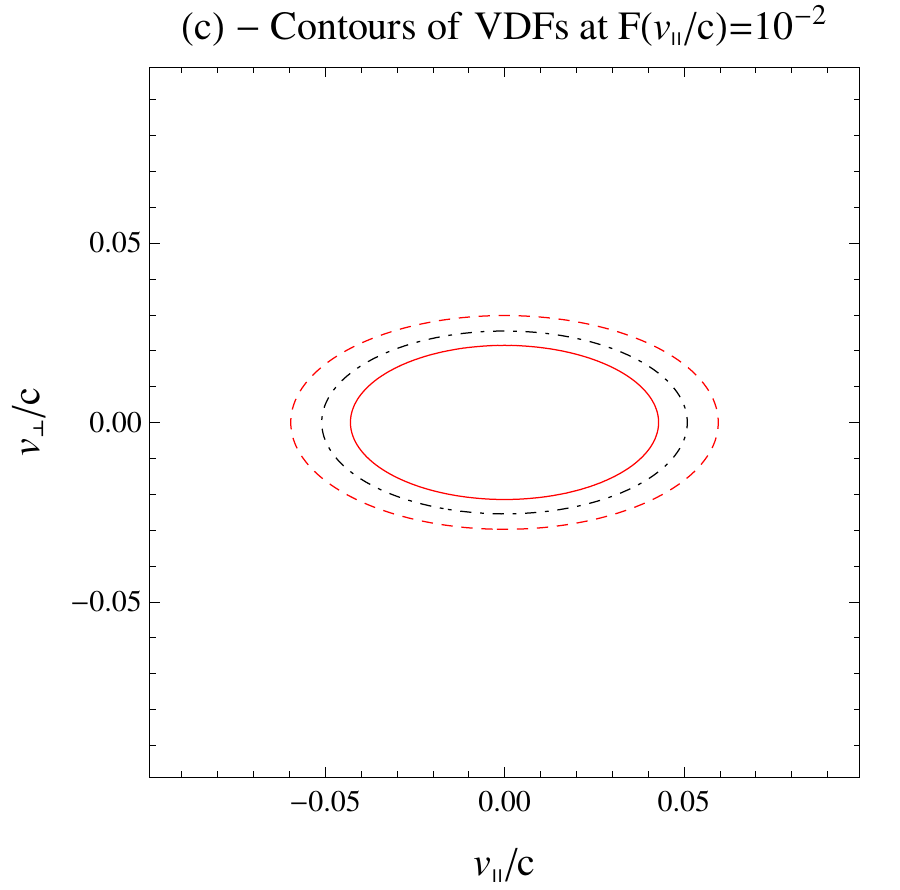}
    \caption{(Color online) Models of VDFs: bi-Kappa from
    Eq.\ (\ref{kappas}) with
    $\theta_{\perp}/c =2 \theta_{\parallel}/c = 0.02$
    (dashed blue lines) and $\kappa = 9/4$, and
    bi-Maxwellian limits ($\kappa \to \infty$) with the same
    temperature (Maxwellian-H with dash-dotted lines)
    or with a lower temperature
    (Maxwellian-C with solid lines). Parallel cuts $F(v_\parallel)$
    are shown in panel (a), and isocontours at 4 10$^{-3}$ in
    panel (b) and 10$^{-2}$ in panel (c), corresponding to
    dotted lines in panel (a). Notice the difference between
    the Maxwellian limits. }\label{fig3}
\end{figure}

We have seen above that not only a $\kappa$-model, as in
Eq. (\ref{kappas}), can be introduced in two different manners
with respect to a given Maxwellian limit by considering the
temperature to be $\kappa$-dependent (Kappa-B) or not (Kappa-A),
but also that both $\kappa$-VDFs can be realized. The difference
in the $\kappa$-VDFs signifies a principal difference of the
corresponding physical systems.

In practice, i.e.\ when interpreting measurements, this contrast
becomes evident in a different way. Suppose that a set of
measurements for a physical system - of which one does not know {\it
a priori} all properties - can be well-fitted by a
$\kappa$-distribution Eq.(\ref{kappas}). It is now, depending on its
interpretation as a Kappa-A or a Kappa-B distribution, possible to
consider two Maxwellian limits, namely, a cooler (C) Maxwellian with
$T^{M,C}_{\parallel,\perp} = (m/k_B) \theta^2_{\parallel,\perp}/2
<T^K_{\parallel,\perp}=(m/k_B)\kappa \theta^2_{\parallel,\perp}/
(2\kappa -3)$, reproducing the low-energy core of the
$\kappa$-VDF, or a Maxwellian limit with a central peak
markedly lower but the same temperature as the $\kappa$-VDF, i.e.\
$T^{M}_{\parallel,\perp}= T^K_{\parallel,\perp}$. For illustration,
such a $\kappa$-VDF and its two Maxwellian limits are shown in
Fig.~\ref{fig3}.

In this situation the above question can be re-phrased into: {\it
Which of the two Maxwellian distributions is the correct limit?}
Again the answer depends on the properties of (or physical processes
realized in) the considered physical system. Relative to the
Maxwellian-C the $\kappa$-distribution shows enhanced high-velocity
tails and a somewhat reduced core, so is of type Kappa-B. This may
enable two distinct applications, namely either to extract the
effects of the suprathermal particles by comparison to the
Maxwellian core (e.g., dissipation and instabilities, as discussed
in \citet{Lazar-etal-2015}) or to model the particle acceleration
\citep{Leubner-2000, Bian-etal-2014}. Alternatively, relative to the
Maxwellian of equal temperature the $\kappa$-distribution exhibits
both enhanced tails and an enhanced core, and is, thus, of type
Kappa-A. This allows to study processes that lead to an accumulation
of particles at low velocities as the one discussed in Sec.~3 above.
The relaxation of a Kappa distribution by keeping the temperature
constant and reducing only the suprathermal tails (eventually
leading to a Maxwellian equilibrium) is also suggested by the
simulations \citep{Vocks-Mann-2003} to be a result of the Coulomb
collisions ($\nu_c \sim v^{-3}$) in the absence of turbulence. This
relaxation seems to ensure the escape of suprathermals from the
corona if their existence is assumed there.

So, for a correct application, one needs to have an idea about the
Maxwellian equilibrium state of the considered system.


\section{Conclusions}

Interestingly, we find that both alternatives for defining
$\kappa$-distributions can be correct, but they are valid for
different physical systems. Kappa-A describes a system in which
a process must exist that enhances the core part of a VDF relative
to its Maxwellian counterpart. While this can be the cause of an
increased effective collision rate provided by wave-particle
interactions, one should expect only specific $\kappa$-values
to be consistent with a given scenario. Kappa-B rather describes
a system, where only a high-velocity enhancement occurs, possibly
due to the lack of sufficient (effective) collisions between the
particles. Thus, Kappa-B appears to be the less specific case and,
thus, should be the more frequently realized alternative. 
With respect to the two alternative Maxwellian limits
of a given $\kappa$-distributed data set it is of significance whether or not an
external source of energy has to be taken into account. The latter case would
correspond to a Kappa-A, the former to a Kappa-B system. 
In any case, before employing one of these two representations,
one should be conscious about the nature of the physical
system one intends to describe in order to avoid to obtain
unphysical results.

\begin{acknowledgements}
The authors acknowledge support from the Katholieke Universiteit Leuven,
the Visiting International Professor (VIP) Programme of the Research
School Plus at the Ruhr-Universit\"at Bochum, and the Deutsche
Forschungsgemeinschaft (DFG) via the grants FI~706/14-1 and
SCHL~201/21-1. P.H.Y. acknowledges the support by the BK21 plus
program through the National Research Foundation (NRF) funded by
the Ministry of Education of Korea. This project has received funding
from the European Union's Seventh Framework Programme for research, technological development
and demonstration under grant agreement SHOCK 284515.
\end{acknowledgements}

\bibliographystyle{aa}
\bibliography{kapint}

\end{document}